\documentstyle[12pt,psfig]{article}
\advance\textheight by\topskip

\begin{document}

\title{
RADIO EMISSION OF THE GALACTIC X-RAYS BINARIES WITH RELATIVISTIC JETS
}
\author{ Sergei A. Trushkin
\and{\it Special Astrophysical Observatory of Russian Academy of Sciences}
\and{\it    357147 Karachaevo-Cherkesia, Nizhnij Arkhyz, Russia }
\and{\it                   E-mail: satr@sao.ru}
}
\date{}
\maketitle

\begin{abstract}

Variable non-thermal radio emission from Galactic X-ray binaries is
a trace of relativistic jets, created near accretion disks.
The spectral characteristics of a lot of radio flares in the
X-ray binaries with jets (RJXB) is discussed in this report.
We carried out several long daily monitoring programs with the
RATAN-600 radio telescope of the sources:
SS433, Cyg X-3, LS\,I $+61^\circ303$, GRS 1915+10 and some others.
We also reviewed some data from the GBI monitoring program at two frequencies
and hard X-ray BATSE (20-100 keV) and soft X-ray RTXE (2-12 keV) ASM data.

We confirmed that flaring radio emission of Cyg X-3 correlated with hard
and anti-correlated with soft X-ray emission during the strong flare
($>3$ Jy) in May 1997.
During two orbital periods we investigated radio light curves
of the remarkable X-binary LS\,I~$+61^\circ303$. Two flaring
events near a phase 0.6 of the 26.5-day orbital period have been
detected for first time at four frequencies simultaneously.

Powerful flaring events of SS433 were detected at six
frequencies in May 1996 and in May 1999.
The decay of the flare is exactly fitted by an exponential law and the
rate of the decay $\tau$ depends upon frequency as $\tau\propto\nu^{-0.4}$ in
the first flare and does not depend upon frequency in the second flare,
and is equal to $\tau=6\pm1$ days at frequencies from 0.96 to 21.7 GHz
in the last flare in May 1999.

Many flaring RJXB show two, exponential and power, laws of flare decay.
Moreover, these different laws could be present in
one or several flares and commonly flare decays are faster at a higher frequency.
The decay law seems to change because of geometric
form of the conical hollow jets. The synchrotron and inverse Compton losses
could explain general frequency dependences in flare evolution.
In conclusion we summarized the general radio properties of RJXB.
\end{abstract}

\subsection*{ X-rays binaries with jets}

The Galactic plane bright X-ray sources are close binaries, in which
a neutron star (NS) or black hole (BH) traps matter from a companion,
forming a accretion disk. Amongst $\sim250$ X-ray binaries only $\sim30$
are detected as radio sources (S$_\nu>1$ mJy). Many (or all!) such bright
sources resolved into the relativistic jets with detectable proper motion
of bright blobs.
This sample of 10-15 such X-ray binaries with jets (RJXB) is extremely
interesting for understanding of the astrophysical relativistic jets
phenomenon.

In this paper I discuss mainly spectral and temporal properties
of the RJXB radio emission.
I would like to address the reader to some remarkable reviews,
devoted to properties of RJXB:
Fender et al. (1997), Mirabel and Rodriguez (1998),
Fender (2000a,b,c), papers from two workshops ``Relativistic Jets
from Galactic Sources'' (1997, Jodrell Bank; 1999, Paris).

Just after the discovery the superluminal sources in 1994,
ten Galactic radio-emitting X-ray binaries
were unified into a sample of RJXB showing relativistic jets.
In such close binaries the accretion disks are created
by flowing of material from a normal star, filling  its Roche lobe,
to a compact one.
The rate of accretion of material in such binaries could change sporadically
or periodically, in accordance with orbital or precession periods.
Then two opposite directed relativistic jets from poles of the accretion disk
around a compact object are created because of a high accretion rate.
Powerful variable radio emission of RJXB seems to form in relativistic
moving matter of jets, and could be a trace of the jets formation in the
Galactic sources. The causes and details of relativistic electrons generation
are unclear, but the accumulated data confirm the strong shocks formation,
spreading within the jets. The circumstellar envelope and strong stellar wind,
forming a hot corona must be added in convincing models.

\begin{figure}[t]
\centerline{\psfig{figure=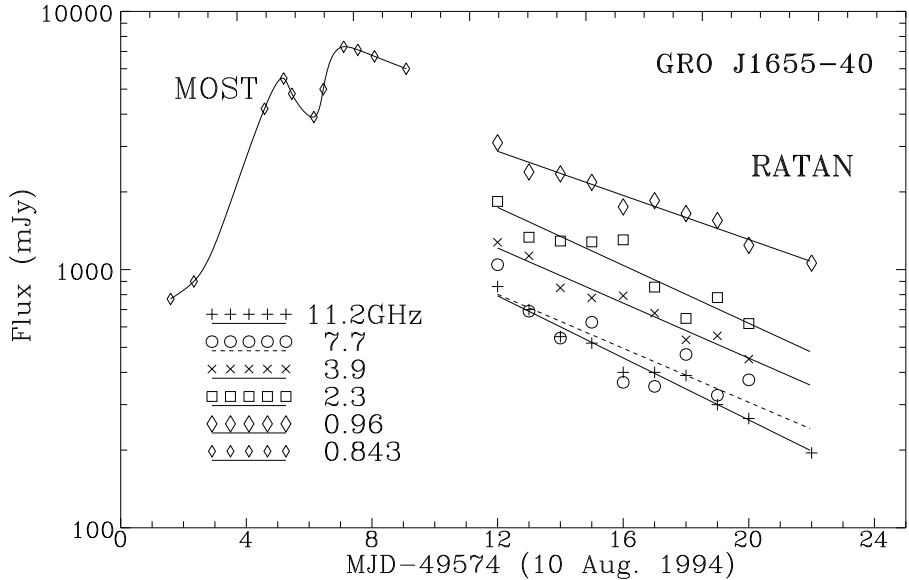,width=12cm,angle=0}}
\caption{%
The flux density variations of GRO J1655$-$40 during power flare in
August 1994 at five frequencies and the MOST data at 843 MHz.
}
\label{16lc}
\end{figure}

The monitoring is continued by Rossi XTE satellite as
All Sky Monitoring (ASM) program during the last 150 weeks.
In radio band Ryle's interferometer at 15 GHz and Green Bank interferometer
(GBI) at 2.25 and 8.3 GHz carried out monitoring of RJXB during some years.
RATAN-600 could obtain complementary daily radio spectra in a wide frequency
range.

\subsection*{The RATAN-600 radio monitoring of RJXB}

In last years we carried out the monitoring program of radio variability
of RJXB: GRO J1655$-$40, LS\,I $+61^\circ303$, GRS 1915+10, SS433, Cyg\,X-1,
Cyg~X-3, and CI Cam.
In 1995--1999 long-time observational sets of RJXB
the dynamical radio spectra at 1--31~cm wavelength range were detected.
We found consistent patterns in spectral variability of RJXB.
Some light curves of the RATAN monitoring  of the RJXB are in
Bursov \& Trushkin (1995), Trushkin (1998) and Trushkin \& Bursov (1999).
The last paper is from the ``Odessa'' part of GMIC'99.

\subsection*{GRO J1655-40}

The dynamically resolved black hole GRO J1655$-$40
is a second (after GRS 1915+105) superluminal X-ray transient discovered
by BATSE in July 1994, with 0.92c jets detected by VLBI and the VLA.
This is bright emission lines variable optical object showing an orbital
modulation of 2.6 days.

\begin{figure}
\centerline{\hbox{
\psfig{figure=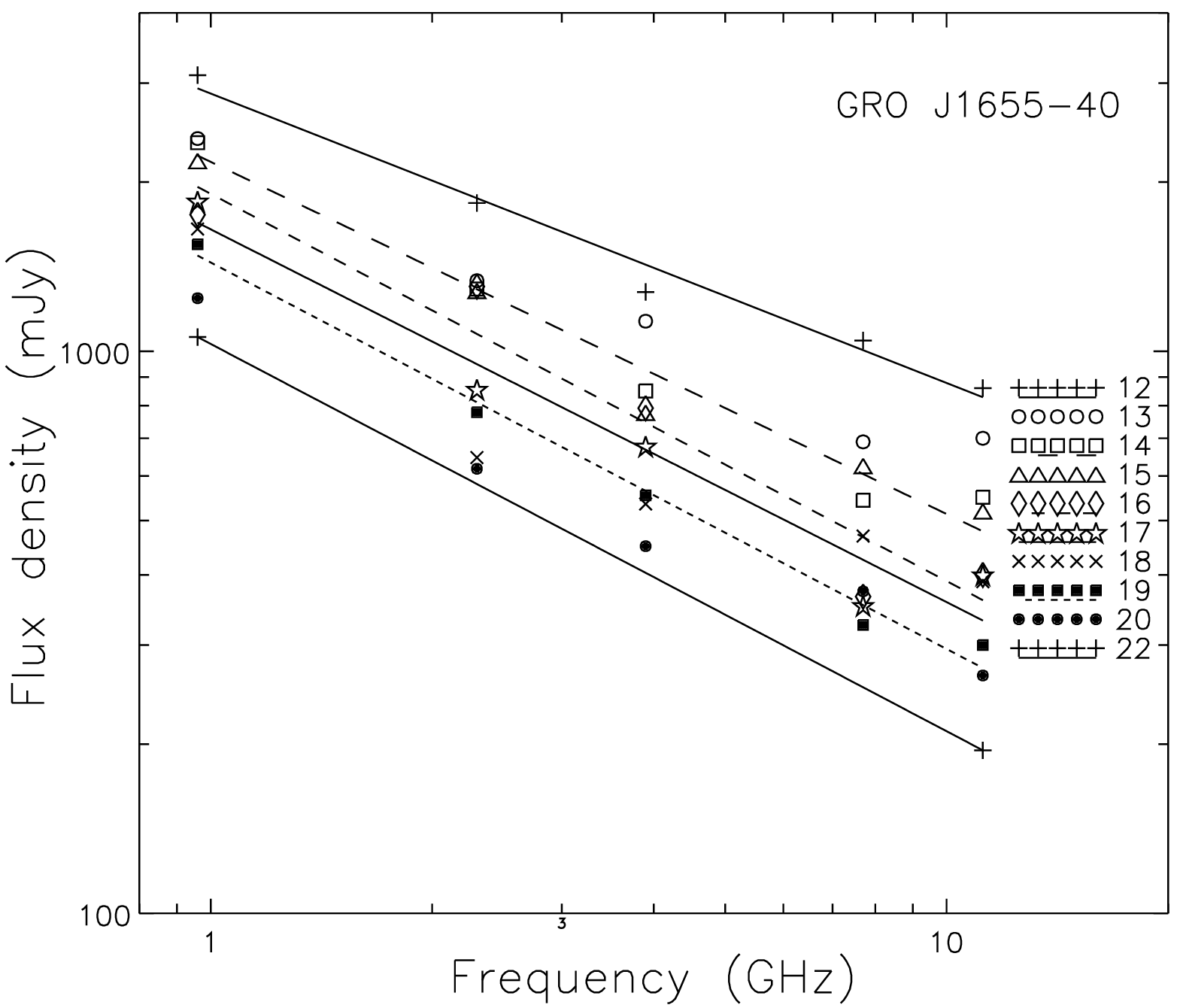,width=8cm,angle=0}
\psfig{figure=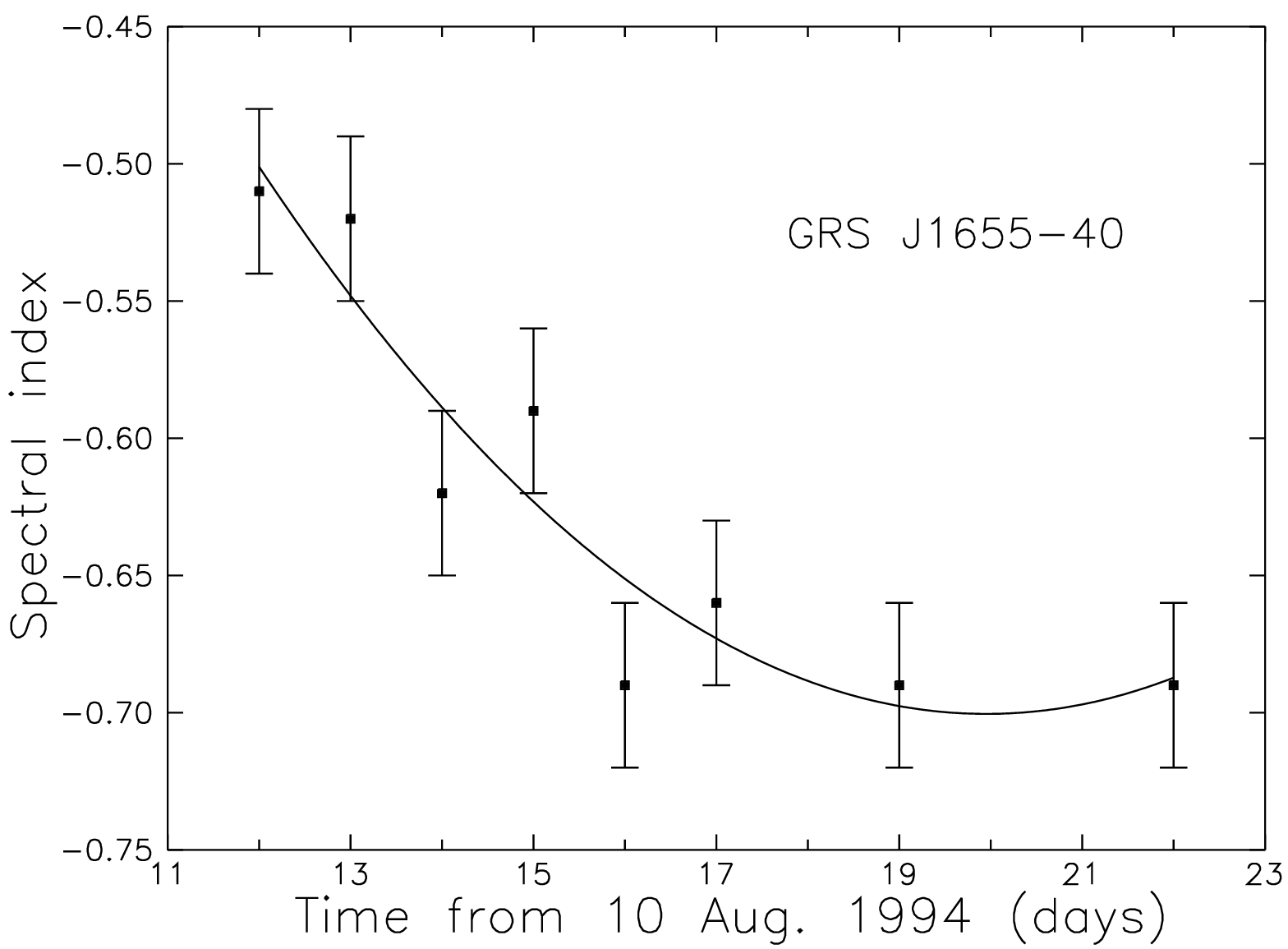,width=8.5cm,angle=0}}}
\caption{%
{\it Left:} The radio spectra GRO J1655$-$40 during the powerful flare in
August 1994 at five frequencies. The figures in legend are days from
the date of the flare beginning (10 August 1994).
{\it Right:} The change of the spectral index of GRO J1655$-$40
during the flare.
}
\label{16sa}
\end{figure}

We observed GRO J1655$-$40 only during the decay of most powerful flare
in August 1994. Fig.\ref{16lc} shows the light curves at five
frequencies and the beginning of the flare from MOST data at 843 MHz.
The straight lines are fits by a exponential law of the flare decay.
The rates of flux decreasing are similar at different frequencies,
but the power law spectra are steeper from 12th to 22th days from the
flare beginning as seen in Fig.\ref{16sa}.

The RATAN data indicate that GRS J1655$-$40 could be in a weak nonthermal
radio shell diameter of $6'$ or 5.6 $d/d_{HI}$ pc
(if distance $d_{HI}=3-6$ kpc) with the
spectrum $S_\nu$[Jy]=0.27$\nu^{-0.5}_{GHz}$. Such a shell does not seem to be
a supernova remnant because it is much weaker than a typical Galactic SNR
with such a surface brightness and diameter.

\subsection*{LSI 61$^\circ$\,303}

\begin{figure}
\centerline{\psfig{figure=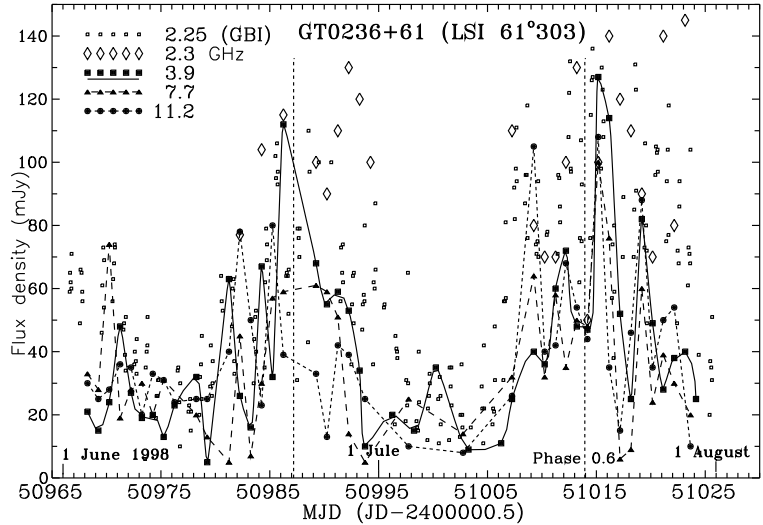,width=10.0cm,angle=0}}
\caption{%
The flux density of LS\,I $61^\circ$\,303 during June-July of 1998
from RATAN and GBI measurements.
Orbital period phases 0.6 are marked by vertical dotted lines.
}
\label{gt}
\end{figure}

The Be star/X-ray source LSI 61$^\circ$\,303 was discovered as the variable
source GT0236+61 in the patrol Galactic plane survey by Gregory and Taylor (1983).
In Fig.\ref{gt} are shown the light curves of LS\,I $61^\circ$\,303
at five frequencies and the GBI data at 2.25 GHz. The x-axis is modified
Julian days. We marked the orbital period phases 0.6, during which usually
the radio flares (or flux maxima) occur again.

Generally the light curves are well correlation,
while the latter maxima of flares come at the lower frequencies.
The delay at 2.3 GHz is equal to 1--2 days
from the maximum of flux at 11.2 GHz.
Daily radio spectra changed quickly day by day and were flat at the
beginning of the flare rise, as a consequence of the delay, and then became
usual optically-thin synchrotron spectra.
Both results could be explained in a model of relativistic jets, moving away
from a compact object in the radio-absorbing dense thermal envelope formed by
a stellar wind of the binary.

The light curves of LSI $61^\circ$\,303 in X-ray band 2--12 keV,
from RXTE satellite (quick-look data in RXTE ASM program; Levine et al.,
1996) show that soft X-ray emission was lower than the detection level during
the radio flux maxima, equal to $\sim5$ mCrab, and just in period the
radio flux minimum (MJD $51000\pm3$) reached a prominent value of 20 mCrab
(0.0362 mJy).

During the observational set we detected variable radio emission
from the recently discovered X-ray binary CI Cam (XTE J0421+560) at 3.9 GHz.
Its nonthermal radio emission was firstly detected at a level of 0.5--1 Jy
at the end of March 1998, just during a powerful X-ray flare.
After three months the radio flux decreased to $\approx50$ mJy and
the source showed small daily flux fluctuations to 20--30\% around
the mean value.

\subsection*{GRS 1915+105}

Mirabel et al. (1994) detected apparent superluminal motions in
first Galactic source GRS 1915+105, X-ray and radio transient mapping of which
with VLA has revealed to possess relativistic jets with
true resolved velocity of 0.92c.

\begin{figure}
\centerline{\psfig{figure=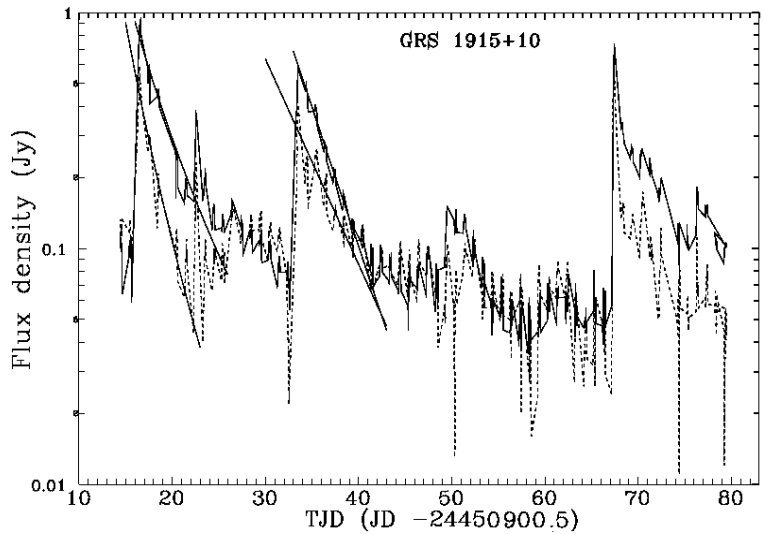,width=10cm,angle=0}}
\caption{%
The flux density variations of GRS 1915+105 during seventy days of 1995
from GBI data at 2.25 GHz and 8.3 GHz.
}
\label{19lc}
\end{figure}

We detected GRS 1915+105 with RATAN-600 observations at 3-4 frequencies
simultaneously. Thus we plotted daily spectra, which  are often inverse with
a positive spectral index of +0.4 -- +0.9 during ``quiet'' periods.
In Fig.\ref{19lc} radio light curves of GRS 1915+105 are shown
at 2.25 and 8.3 GHz (JD = 2450900.5 is 00UT 28 March 1998).
We used a semi-log plot in order to show the laws of flares decay.
Unfortunately it is difficult to say which law is present really.
The first and second powerful flares are best fitted by power and exponential
laws, respectively while a last law is better fit in general.
The spectral index changed from $-0.4$ to $-1.0$ for the first
flare in Fig.\ref{19lc} and, in contrast, for second one from
$-0.2$ to $+0.5$.

Possible association GRS 1915+105 with SNR G45.7$-$0.4 is unclear.
The X-ray binary is located at the south-west boundary of SNR.
The defined size of this SNR ($22'$) is very indefinite and G45.7$-$0.4
could be wider in X-ray binary direction, as seen from the NVSS map
at 1.4 GHz. The distances to SNR and
GRS 1915+105 are comparable: d($\Sigma-D$)=9-10 kpc, and d(GRS)=10-12.5 kpc.

\subsection*{Cyg X-3}

Since 1972, when Gregory detected a powerful variable radio source,
associated with the X-ray binary Cyg X-3, it is regular monitored at
radio frequencies. Trushkin (1998) described in details some interesting
flaring events during the 80-days monitoring set in summer of 1997.
It is coincided with collaborating multi-band monitoring program.
Analysis of the light curves in X-ray and radio range (McCollough et al.,
1998a,b) confirmed a high correlation of the hard X-ray flux
(20--100 keV) and flaring radio flux, and a anticorrelation
with soft X-ray emission (2--12 keV) during powerful flaring activity.
These light curves are shown by Trushkin \& Bursov (1999).
For the powerful flare dates of the correlation coefficients are found:
$\rho$ (RXTE  -- 11 GHz) =  $-0.64\pm0.04$,
$\rho$ (BATSE -- 11 GHz) =  $0.54\pm0.04$.
Then for the post-flare period the picture is changed:
$\rho$ (RXTE  -- 3.9 GHz) =  $+0.69\pm0.01$  or
$\rho$ (RXTE -- 11 GHz) =  $+0.66\pm0.02$.
In general the soft and hard X-ray emission show the anticorrelation in total
active period: $\rho$ (2-10 keV -- 20-100 keV) =  $-0.64\pm0.01$,
that also confirms before discovered dependence, that a hardness of the
X-ray emission anti-correlates with its brightness in a flare.

Probably a flaring variability is a direct evidence of jets formation
and their expanding in thermal shell around Cyg X-3 (Trushkin, 1998).
On other hand during the deep minimum of the radio flux
(below 10 mJy) the formation of the jets are ceased temporarily, in 1--2
weeks. And we see only weak radio emission with flat spectrum from envelope.

The variability of Cyg X-1,  the well-known black hole candidate,
was studied at 3.9 GHz during the same set.
Non-thermal radio emission weakly fluctuated in a range from 10 to 30 mJy.
We could not find any significant radio flux modulation
with orbital period 5.6 days,
recently detected at high frequencies (Pooley et al., 1999).
The orbital modulation of the soft X-ray emission was in detail investigated
using two-year monitoring program data with RTXE (Wen et al., 1999).

In the 40-day set of observations in December 1998 -- January 1999
flux variability of the Cyg X-3 is monitored during
very low soft X-ray flux ($\sim80$ mCrab in range  2-12 keV).
Then optically thick radio emission has positive spectral index ($\sim+0.3$)
in the range 2.3--11.2 GHz.
A flux changed from 40 to 160 mJy at the frequencies.

\subsection*{SS433}

The X-ray binary and luminous star SS 433 is a persistent bright radio source,
detected first in 4C survey. Its active periods are characterized by
powerful flares with the flux increasing two-ten times
during 1-2 days (e.g. Bursov \& Trushkin, 1995).
SS433 was resolved into radio blobs on scales from $0.005''$ to $3''$,
the radio structure location followed a kinematic model, constructed from
``moving'' emission lines, originated in two opposite directing precessing
jets.

\begin{figure}
\centerline{\hbox{
\psfig{figure=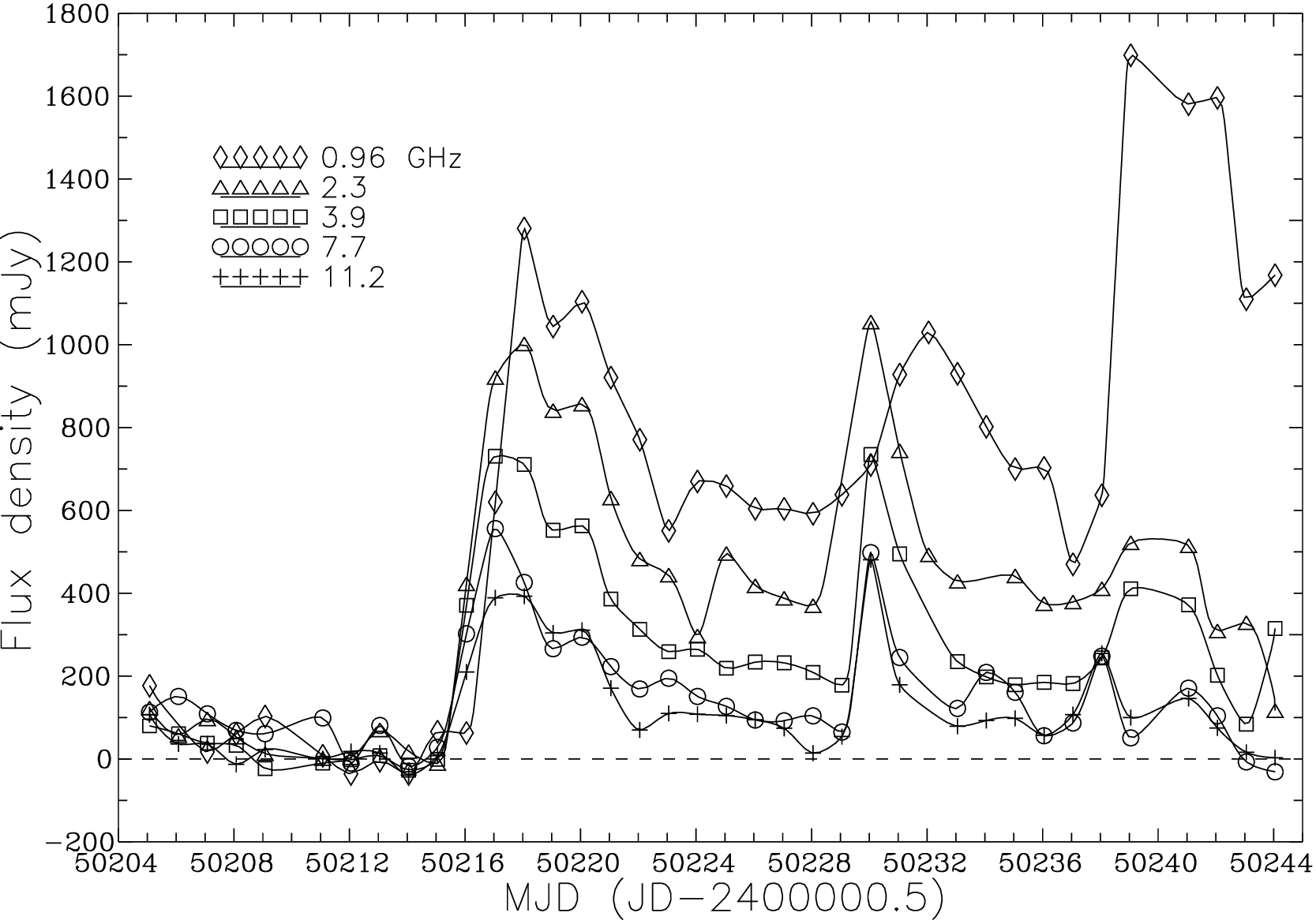,width=8.2cm,angle=0}
\psfig{figure=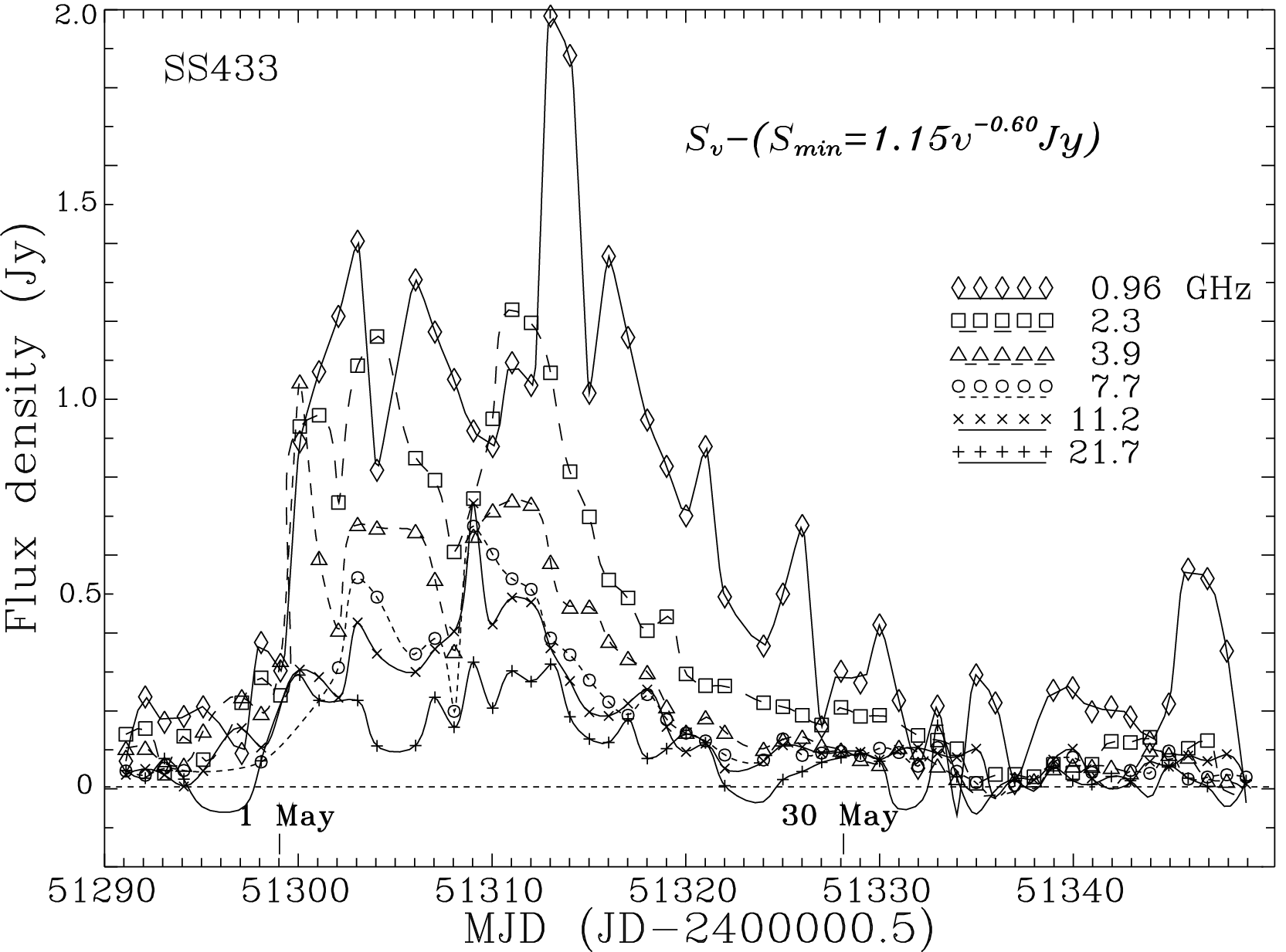,width=8.2cm,angle=0}}}
\caption{%
Light curves of SS433 at five-six frequencies in May-June 1996 ({\it left})
and May-June 1999 ({\it right}). The quiet spectrum was subtracted.
}
\label{SS9699}
\end{figure}

\begin{figure}
\centerline{\hbox{
\psfig{figure=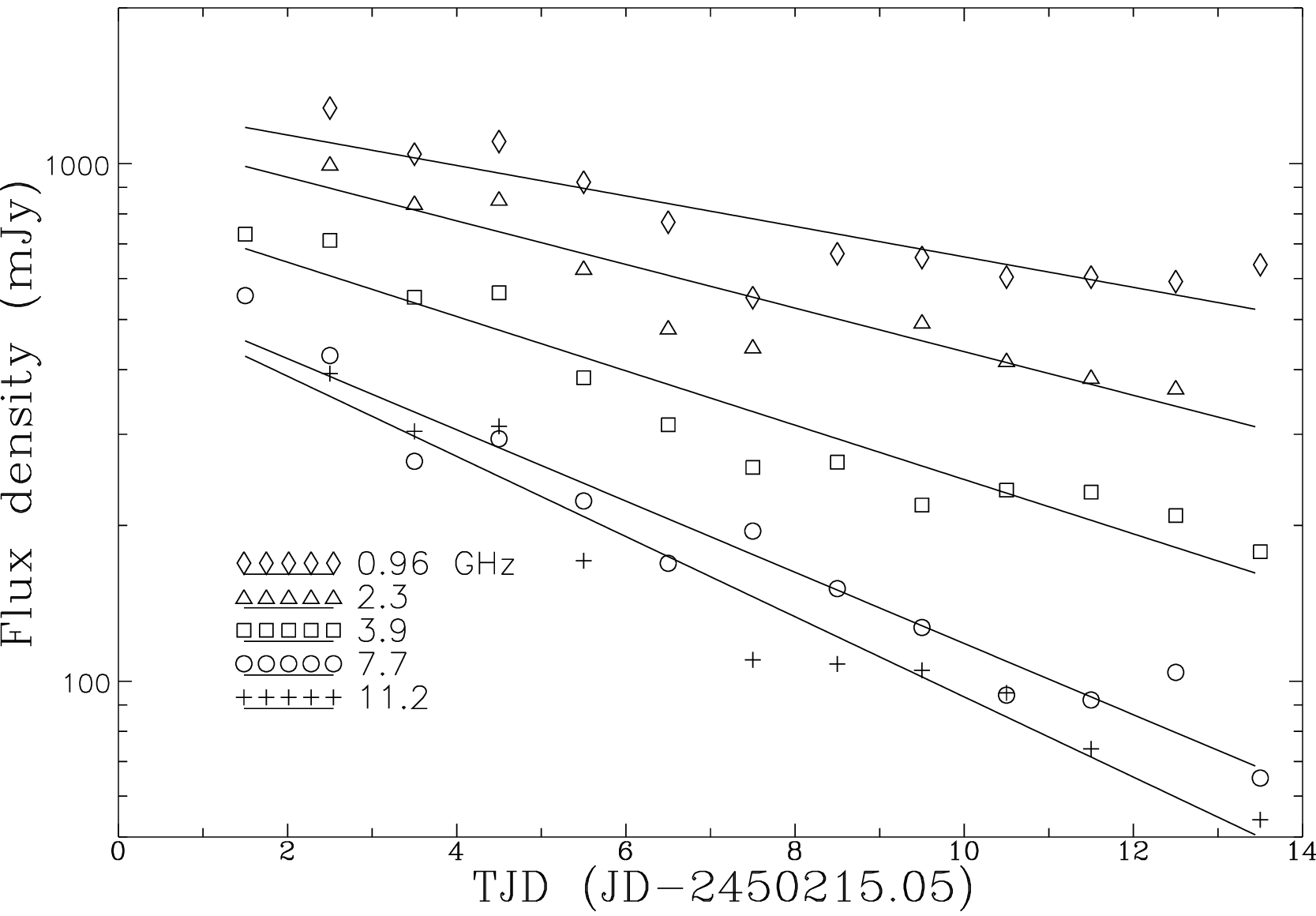,width=8cm,angle=0}
\psfig{figure=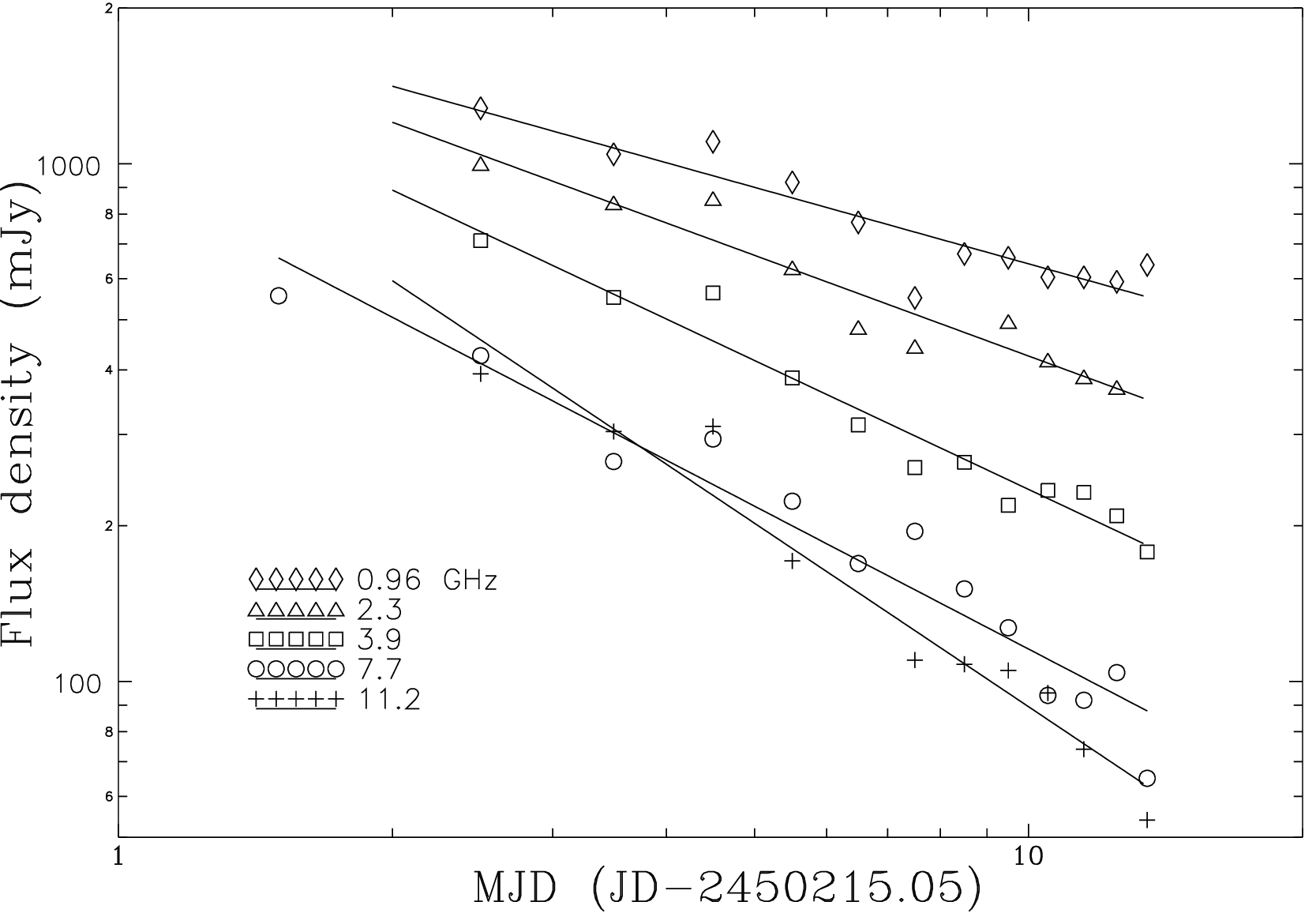,width=8cm,angle=0} } }
\caption{Exponential and power laws fittings
for the same SS433 flare in May 1996.
}
\label{SSf1}
\end{figure}

Here we can discuss only some characteristics of the powerful
flares of SS433 in the last years.

In Fig.\ref{SS9699} the radio light curves of the SS433 flares, derived
with the continuum radiometric complex of the RATAN-600 radio telescope
are given.
The usual quiet spectrum, defined in the date range of MJD50210--50215
was subtracted.

\begin{figure}
\centerline{\hbox{
\psfig{figure=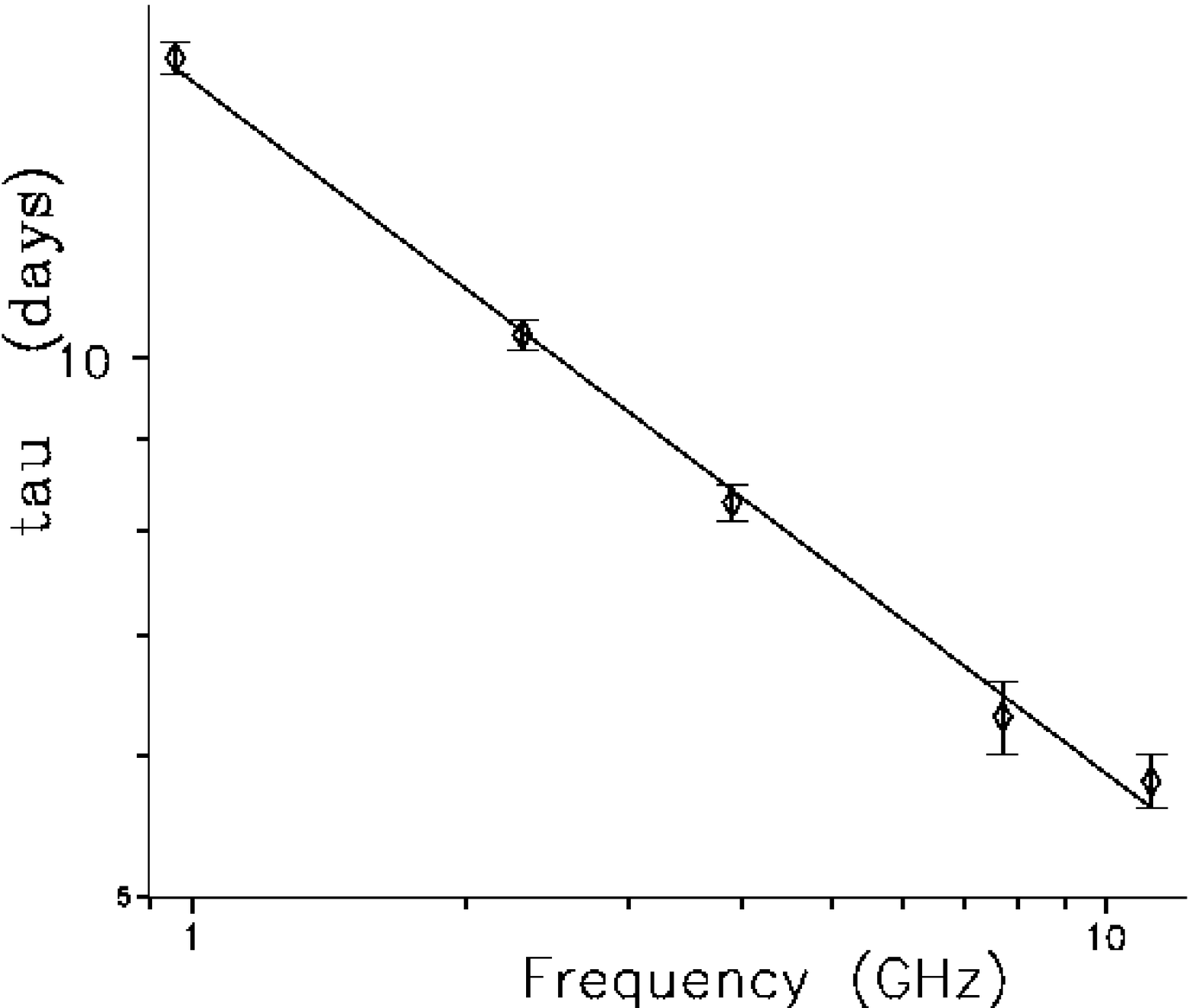,width=8cm,angle=0}
\psfig{figure=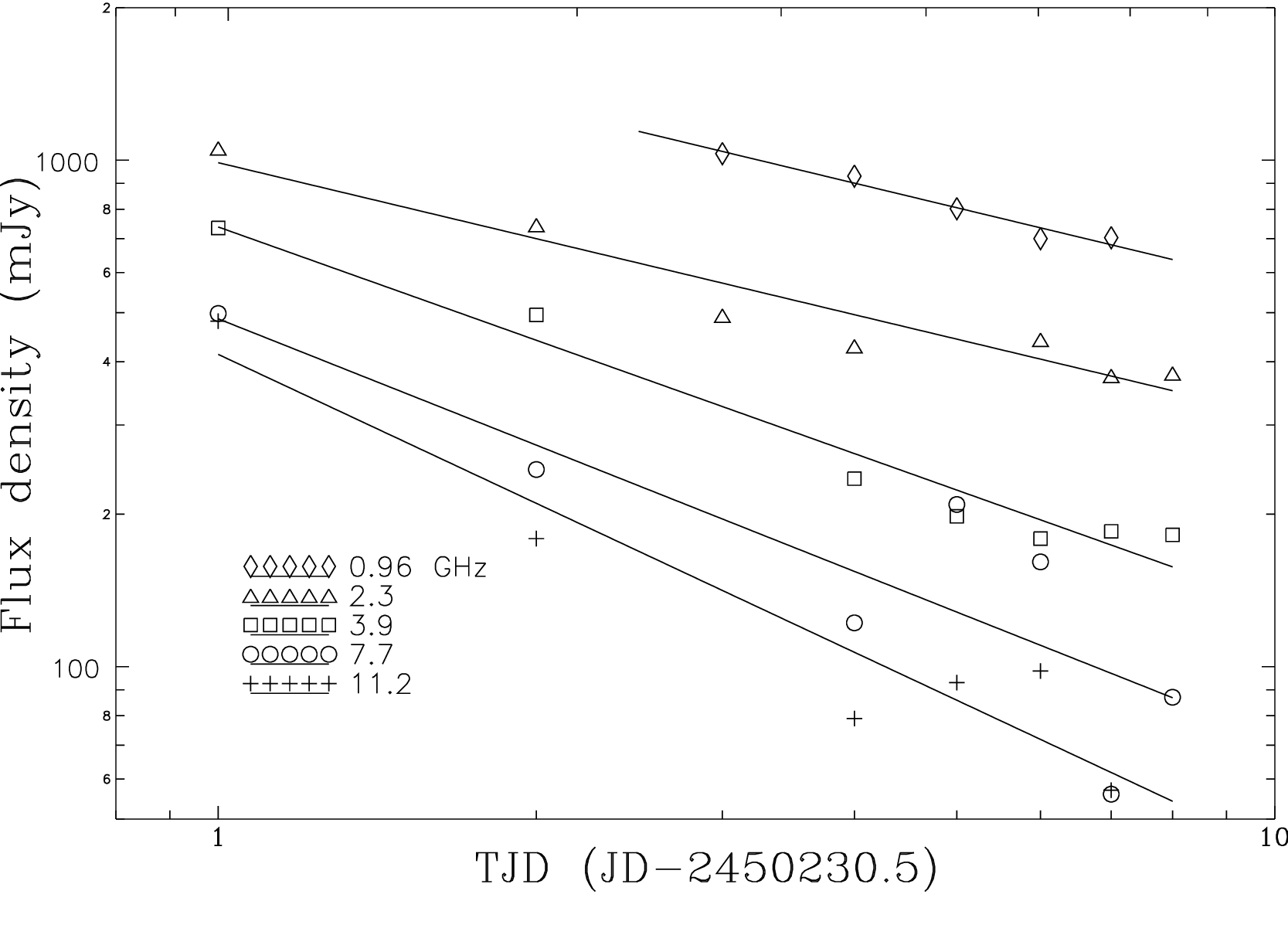,width=7cm,angle=0} } }
\caption{%
{\it Left:} Decay rates ($\tau$) of the first SS433 flare in May 1996.
defined for exponential fitting of the decay via frequency.
{\it Right:} Power law fittings of the second flare decay in May 1996}
\label{SSt2}
\end{figure}

In Fig.\ref{SSf1} plots of the first flare decays of SS433 are given
on different log/semi-log scales with two different fits for comparison.
In the figures there are no a distinguished difference of the fits,
but the exponential law gives other important and reliable dependence
--  a power law frequency dependence of a decay rate.
In Fig.\ref{SSt2} (left) this dependence is shown. The fitting gives
$\tau$(days)=14.3\,$\nu^{-0.4}_{GHz}$. It allowed me to argue
that the flare decay follows the exponential law.
In Fig.\ref{SSt2} (right) the best power law fitting of the second flare
in May 1996 is shown, indicating that these dependences are steeper
with increasing frequency, from -0.5 to -1.0 at 0.96 and 11.2 GHz,
respectively.

In the 60-day set of observations in April 23 -- June 23 1999
the strong flare of SS433 was detected during relative low soft
X-ray emission state -- 20 mCrab in the range 2--12 keV.
The data of GBI monitoring of SS433 show that its X-ray and radio activity
increase in last year. That is determined by common increasing quiet level
in 1.5 times at all frequencies with constant spectral index
($\sim-0.6$) and increasing the frequency of powerful X-ray and radio flares.
(see Trushkin \& Bursov, 1999).

\begin{figure}
\centerline{\psfig{figure=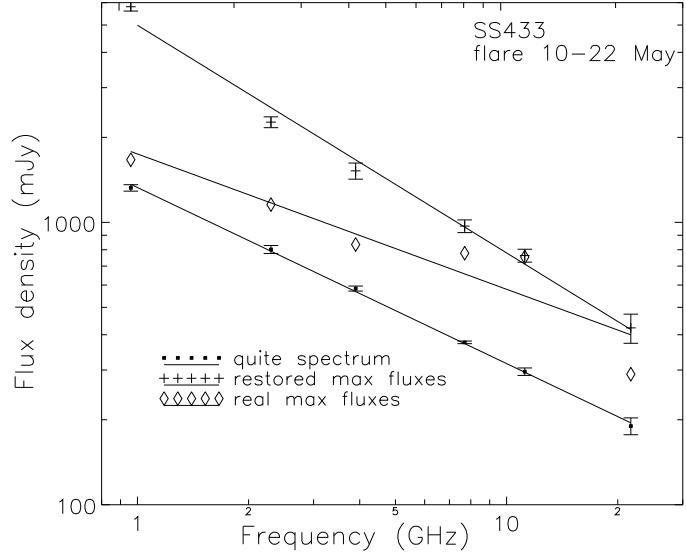,width=9.cm,angle=0}}
\caption{``Restored'' spectrum at the beginning of the SS433 flare of
May 10--20 1999.
Its spectral index is equal to $-0.8$, while the spectra of real maximum flux
densities have a spectral index equal to $-0.40$.}
\label{SSfm}
\end{figure}

In Fig.\ref{SS9699}(right) radio light curves of the SS433 flares
detected in May 1999.
We subtracted the ``minimum quiet state'' that is determined
far long ego and being coincident to fluxes in the beginning of June 1999:
$S_\nu$[Jy] = $1.15\,\nu^{-0.60}_{\mathrm GHz}$.
Daily radio spectra during the set indicated a negative spectral index.
The flare spectrum became steeper during its decay.

A delay of flaring maxima increase with decreasing of the frequency,
and is particularly traceable at 0.96 GHz.
Decline of the flare (May 10--28) is exactly fitted
by exponential law at all frequencies, that is reliable
during twenty days at different frequencies. In contrast with the first flare
in May 1996 there are no any detectable dependence of rate of decreasing
$\tau$ ($S_\nu \sim S_\circ\,exp(-t/\tau)$) upon frequency.
This rate of decay seems to be similar $\tau=6\pm1$ days
at six frequencies.

If a light curve in the beginning of a flare
are characterized by a high absorption in a thermal circumstellar envelope,
then a spectrum of $S_\circ$ does indeed indicate the initially injected
distribution of the relativistic electrons.
In Fig.\ref{SSfm}
dependence $S_\circ$ on frequency is exactly followed to a power law
with spectral index equal to $-0.8$ that is usual for non-thermal
cosmic sources, like quasars or radio galaxies,
indicating acceleration of relativistic electrons on strong shocks in jets.

\begin{figure}
\centerline{\psfig{figure=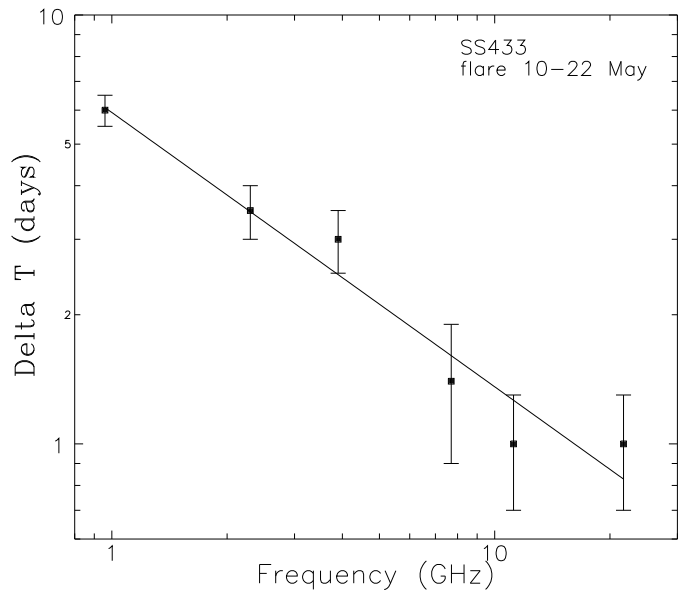,width=9.cm,angle=0}}
\caption{ Shifts $\Delta\,T$ of maximum fluxes at different frequencies from
the beginning of the SS433 flare of May 10--20 1999 via frequency.
It well enough fitted by a power law with index of $-0.64$. }
\label{SSdt}
\end{figure}

In Fig.\ref{SSdt}
shifts of maximum fluxes dependence, $\Delta\,T$ on frequency are given.
It is exactly followed to a power law with spectral index equal to
$-0.64$ that is usual for non-thermal cosmic sources, like quasars or radio
galaxies,
indicating acceleration of relativistic electrons on strong shocks in jets.
It is commonly for SS433 that $\Delta\,T(\nu)$ is exactly followed to a
power law with spectral index being in range from $-$0.8 to $-$0.4.

The common property of RJXB is anticorrelation a radio flaring radio flux
and a level of soft X-ray emission in comparison of these data.
Usually X-ray flares coincided with
quiet periods in radio light curves and, vice versa,
rare and powerful radio flares coincided with periods of low X-ray emission,
probably formed in far regions around SS433.
The variable hard X-ray emission, originated from inverse Compton scattering
of stellar photons off relativistic electrons responsible for the radio flux
could strongly increase during flaring events.

\subsection*{Conclusions}
Below we summarize some general radio emission parameters in RJXB:

$\bullet$
Sources Cyg\,X-3, Cir\,X-1, CI Cam (XTE 0421+560), 1E1740.7$-$2942,
GRS 1758$-$258, GX339$-$4, LSI+$61^\circ303$ and SS433, are radio jet X-ray binaries.
Probably jets play a key role in  formation of powerful non-thermal
radio emission. The correlation of hard X-ray and radio emission
seems to be a common feature of RJXB.

$\bullet$ All RJXB are strongly variable X-ray and IR sources.
The Cir X-1 show the neutron star and GRO J1655$-$40 is a dynamically resolved
system with a black hole. Other binaries are only probable NS or BH.

$\bullet$ VLBI observations of RJXB show multi-component structure
on a scale $0.001-5''$. High velocities, $0.1-0.92$c, are detected
from the proper motion of blobs in resolved sources, often showing a
apparent superluminal expansion.
The relativistic electrons and magnetic fields reserved a large portion
of the total power of flare.

$\bullet$ The synchrotron spectra $S_\nu=S_\circ\nu^\alpha$
of RJXB are variable with flares up to 1000 times comparing with quiescent
state.
Also high linear polarization was detected in SS433, Cyg X-3 during the flares.
The processes of thermal absorption in a dense envelope are used for explanation
of the frequency-dependent delays of flare maxima.

$\bullet$ The basic model of the synchrotron emission evolution is
an adiabatic expansion of the blobs moving away from binaries, which contain
the relativistic electrons and magnetic fields (Shklovski, 1960,
van der Laan, 1966; then Marti et al., 1992).
A conical geometry of jets and considerations of the
radiative losses, synchrotron radiation and inverse Compton scattering (ICS)
are a modification the basic model to satisfy the spectral and temporal
dependences during the flare evolution.
ICS in the hot corona around a binary
could be responsible for correlation of hard X-ray and radio emission.

$\bullet$ Monitoring of radio variability shows that the decay of flaring
flux after the maximum follows:
a power law $S_\nu=S_\circ\,t^{-2p}$, as the Shklovski-van der Laan model
predicts and a exponential law $S_\nu=S_\circ\,e^{-t/\tau}$
the cause of which is a geometric structure of jets.
Here often (but not always) $\tau \sim \nu^\beta$,
where $\beta$ ranged from $-$0.8 to $-$0.4, thus a flare decays
faster at the higher frequency.
Maybe there are two different types of flare in
RJXB with or without delays and different laws of decay: ``flare of core'' and
``brightening zone'' in SS433.

$\bullet$ Quasi-periodical oscillations are detected at X-ray
or radio light curves with characteristic frequencies: 0.01-1000 Hz in
the black hole candidate RJXB.

$\bullet$ Non-thermal radio halos are produced by precessing jets.
Good examples are SS433 and Cir X-1, probably GRS J1655$-$40 could be
in a weak radio shell, and GRS 1915+10 seems to be associated with SNR.

$\bullet$ Similar (magnetic, hydrodynamic, relativistic) models of jet
confinement are applied for the Galactic and extragalactic jets despite
their different spatial scales.

{\bf Acknowledgements}.
{\small Author is thankful to RFBR for supporting the project
of monitoring X-ray binaries, grant N98-02-17577.}

\end{document}